\documentclass[aps,pra,twocolumn,superscriptaddress]{revtex4}
\usepackage{bm}
\usepackage{mathrsfs}
\usepackage{amsmath}
\usepackage{amssymb}
\usepackage{graphicx}
\usepackage{amsfonts}
\usepackage{amsthm}
\usepackage{color}
\usepackage{dcolumn}
\usepackage{txfonts}
\usepackage{epsfig}
\usepackage{enumitem}

\begin{document}
\title{Linking Phase Transitions and Quantum Entanglement at Arbitrary Temperature}
\author{Bo-Bo Wei}
\email{bbwei@szu.edu.cn}
\affiliation{School of Physics and Energy, Shenzhen University, 518060 Shenzhen, China}

\begin{abstract}
In this work, we establish a general theory of phase transitions and quantum entanglement in the equilibrium state at arbitrary temperatures. First, we derived a set of universal functional relations between the matrix elements of two-body reduced density matrix of the canonical density matrix and the Helmholtz free energy of the equilibrium state, which implies that the Helmholtz free energy and its derivatives are directly related to entanglement measures because any entanglement measures are defined as a function of the reduced density matrix. Then we show that the first order phase transitions are signaled by the matrix elements of reduced density matrix while the second order phase transitions are witnessed by the first derivatives of the reduced density matrix elements. Near second order phase transition point, we show that the first derivative of the reduced density matrix elements present universal scaling behaviors. Finally we establish a theorem which connects the phase transitions and entanglement at arbitrary temperatures. Our general results are demonstrated in an experimentally relevant many-body spin model.
\end{abstract}
\pacs{05.70.-a, 03.65.Yz,05.70.Ln}
\maketitle

\section{Introduction}
Quantum phase transition is a transition between different quantum phases of a many-body system at zero temperature \cite{Sachdev2011,Cardy1996}.
It comes from diverging quantum fluctuations and may be observed by varying the control parameter of the system at zero temperature \cite{Sachdev2011}. In recent years, a large amount of effort has been made in investigating phase transitions from the perspective of quantum information science \cite{QI2000}, in particular the quantum entanglement \cite{entanglementQPT2002,entanglementQPT2003,entanglementQPT2008,Renyientanglement2010}  and the quantum fidelity \cite{Sun2006,Zanardi2006,GuReview}. The advantage of investigating phase transitions from quantum information science approach compared to the conventional approach is that one do not need to know the local order parameter of the phase transitions and specific symmetries of microscopic Hamiltonian \cite{You2007,Zanardi2007,Venuti2007,YangMF2007,YangMF2008,Paun2008,Chen2008,Gu2008a,Gu2008b,Gu2008c,Gu2009,Schwandt2009,fscaling2010,fs2011,fs2012a,
fs2012b,fs2012c,fs2013a,fs2013b,fs2013c,fs2014a,fs2014b,You2015,fs2015a,fs2015b,fs2017,fs2017b}.

Previous investigations on the relations between phase transitions and entanglement are based primarily on specific many-body models \cite{entanglementQPT2002,entanglementQPT2003,entanglementQPT2008,Renyientanglement2010}. Recently, Wu and his collaborators \cite{Wu2004,Wu2005} studied the relations between quantum phase transitions and quantum entanglement in a general settings and their theories are valid for a very broad class of many-body systems \cite{Wu2004,Wu2005,Wei2016NJP,Wei2017SR}. However, Wu's results are valid only at zero temperatures. Realistic experiments are performed at nonzero temperatures. It is thus highly desirable to investigate whether the general relations between entanglement and phase transitions survive at nonzero temperature.

Motivated by the works of Wu and his collaborators \cite{Wu2004,Wu2005}, in the present work, we study the general relations of phase transitions and entanglement in the equilibrium state at arbitrary temperatures. We derived a set of universal functional relations between the matrix elements of two-body reduced density matrix of the canonical equilibrium state and the Helmholtz free energy of the equilibrium state. This reveals that the Helmholtz free energy and its derivatives are directly related to entanglement measures since any entanglement measures are defined from the reduced density matrix. We show that the first order phase transitions are signaled by the matrix elements of reduced density matrix while the second order phase transitions are witnessed by the first derivatives of the reduced density matrix elements. Close to second order phase transition point, we show that the first derivatives of the reduced density matrix elements present universal scaling behaviors. Finally we establish a theorem which connects the phase transitions and entanglement at arbitrary temperatures. We demonstrated our general conclusions in the Lipkin-Meshkov-Glick (LMG) model which presents both quantum phase transitions and thermal phase transitions.

This paper is structured as follows. In Sec.~II, we establish the general framework and derived the relations between Helmholtz free energy and the reduced density matrix elements. In Sec.~III, we establish the relations between
phase transitions and reduced density matrix. Sec.~IV is devoted to study the relations between phase transitions and entanglement. In Sec.~V, we study the LMG model to demonstrate our general results. Finally Sec.~VI is a brief summary and discussion.

\section{Free Energy and Reduced Density Matrix}
Let us consider a general Hamiltonian up to two-body interactions,
\begin{eqnarray}\label{Ham}
\mathcal{H}=\sum_{i,\alpha,\beta}\mathcal{E}_{\alpha\beta}^i|\alpha_i\rangle\langle\beta_i|+\sum_{i,j,\alpha,\beta,\gamma,\delta}\mathcal{V}_{\alpha\beta\gamma\delta}^{i,j}|\alpha_i\beta_j\rangle\langle\gamma_i\delta_j|.
\end{eqnarray}
Here $\{|\alpha_i\rangle\}$ is a complete basis for the Hilbert space and $\alpha,\beta,\gamma,\delta\in[0,1,\cdots,d-1]$ with $d$ being the dimension of the Hilbert space and $i,j$ are the indices labelling $d$-level systems (qudits).
This Hamiltonian is the same as that discussed in \cite{Wu2004} where quantum phase transitions and reduced density matrix are discussed. In the present work, we generalize the connections between phase transitions and entanglement at zero temperature to arbitrary temperatures. At non-zero temperature,
the canonical density matrix of a many-body system with Hamiltonian $\mathcal{H}$ which is in thermal equilibrium with a heat bath at fixed temperature $T$ is given by
\begin{eqnarray}\label{dm1}
\rho=\frac{e^{-\beta\mathcal{H}}}{Z},
\end{eqnarray}
where $\beta=1/T$ is the inverse temperature of the bath (We set the Boltzmann constant $k_B=1$) and $Z=\text{Tr}[e^{-\beta\mathcal{H}}]$ is the canonical partition function of the system.
From the canonical density matrix \eqref{dm1}, the Helmholtz free energy is thus given by
\begin{eqnarray}
F&=&E-TS,\label{F1}\\
&=&\text{Tr}[\rho\mathcal{H}]+T\text{Tr}[\rho\ln\rho],\label{free}\\
&=&\sum_{i,\alpha,\beta}\mathcal{E}_{\alpha\beta}^i\text{Tr}\left[\rho|\alpha_i\rangle\langle\beta_i|\right]
+\sum_{i,j,\alpha,\beta,\gamma,\delta}\mathcal{V}_{\alpha\beta\gamma\delta}^{i,j}\text{Tr}\left[\rho|\alpha_i\beta_j\rangle\langle\gamma_i\delta_j|\right]\nonumber\\
&&+T\text{Tr}[\rho\ln\rho],\\
&=&\sum_{i,\alpha,\beta}\mathcal{E}_{\alpha\beta}^i\rho_{\alpha\beta}^i+\sum_{i,j,\alpha,\beta,\gamma,\delta}\mathcal{V}_{\alpha\beta\gamma\delta}^{i,j}\rho_{\gamma\delta,\alpha\beta}^{ij}+T\text{Tr}[\rho\ln\rho],\\
&=&\sum_{ij}\text{Tr}[\mathcal{U}^{ij}\rho^{ij}]+T\text{Tr}[\rho\ln\rho]\label{F7}.
\end{eqnarray}
Here in Equation \eqref{F1}, $E$ is the internal energy and $S$ is the entropy. In Equation \eqref{F7}, $\rho^{ij}$ is two-body reduced density matrix of the canonical density matrix $\rho$ and $\mathcal{U}^{ij}$ is defined by
\begin{eqnarray}
\mathcal{U}_{\alpha\beta,\gamma\delta}^{ij}&=&\mathcal{E}_{\alpha\gamma}^i\delta_{\beta\delta}^j/\mathcal{N}_i+\mathcal{V}_{\alpha\beta\gamma\delta}^{i,j}.
\end{eqnarray}
Here $\mathcal{N}_i$ is the number of qudits that the qudit $i$ interact with and $\delta_{\beta\delta}^j$ is the Kronecker delta function on the qudit $j$.
The internal energy of the equilibrium state can be given by
\begin{eqnarray}\label{internal}
E&=&\text{Tr}[\rho\mathcal{H}]=\sum_{ij}\text{Tr}[\mathcal{U}^{ij}\rho^{ij}].
\end{eqnarray}
The functional relations in Equation \eqref{F7} and \eqref{internal} tell us that the internal energy of the equilibrium state is fully determined by the two-body reduced density matrix of the canonical density matrix but the free energy do not. Equation \eqref{internal} not only holds for the internal energy but also for other physical observable. For example the average value of an arbitrary two-body operator $\mathcal{M}=\sum_{ij}\mathcal{M}^{ij}$ in the equilibrium state is fully characterized by the two body reduced density matrix as
\begin{eqnarray}\label{manybody}
\langle \mathcal{M}\rangle&=&\sum_{ij}\text{Tr}\left[\rho^{ij}\mathcal{M}^{ij}\right].
\end{eqnarray}
Note that Equation \eqref{internal} and Equation \eqref{manybody} can be easily generalized to Hamiltonian with $n$-body interactions, where the internal energy and the average value of any physical observable are connected to $n$-body reduced density matrix.

\section{Phase Transitions and the Reduced Density Matrix}
In this section, we establish the relations between phase transitions and the reduced density matrix. We assume the many-body Hamiltonian $\mathcal{H}$ in Equation \eqref{Ham} depends on control parameter $\lambda$ through $\mathcal{E}_{\alpha\beta}^i$ and $\mathcal{V}_{\alpha\beta\gamma\delta}^{i,j}$. Assuming that $\mathcal{E}_{\alpha\beta}^i$ and $\mathcal{V}_{\alpha\beta\gamma\delta}^{i,j}$ are smooth functions of the control parameter of the system $\lambda$. From the definition of free energy, Equation \eqref{free} we have
\begin{eqnarray}
\frac{\partial F}{\partial\lambda}&=&\text{Tr}\left[\frac{\partial\rho}{\partial\lambda}\mathcal{H}+\rho \frac{\partial\mathcal{H}}{\partial\lambda}\right]
+T\text{Tr}\left[\frac{\partial\rho}{\partial\lambda}\ln\rho+\rho \frac{\partial\ln\rho}{\partial\lambda}\right],\label{f1}\\
&=&\text{Tr}\left[\frac{\partial\rho}{\partial\lambda}\mathcal{H}+\rho \frac{\partial\mathcal{H}}{\partial\lambda}\right]
+T\text{Tr}\left[\frac{\partial\rho}{\partial\lambda}\ln\rho\right],\label{f2}\\
&=&\text{Tr}\left[\frac{\partial\rho}{\partial\lambda}\mathcal{H}+\rho \frac{\partial\mathcal{H}}{\partial\lambda}\right]
-T\text{Tr}\left[\frac{\partial\rho}{\partial\lambda}(\beta\mathcal{H}+\ln Z)\right],\label{f3}\\
&=&\text{Tr}\left[\rho \frac{\partial\mathcal{H}}{\partial\lambda}\right],\label{f4}\\
&=&\sum_{ij}\text{Tr}\left[\frac{\partial\mathcal{U}^{ij}}{\partial\lambda}\rho^{ij}\right]\label{f5}.
\end{eqnarray}
In the above derivation, the last term in Equation \eqref{f1} which reduces to $\text{Tr}[\partial_{\lambda}\rho]=0$ vanishes because of the normalization condition of the density matrix $\text{Tr}[\rho]=1$.
From Equation \eqref{f2} to Equation \eqref{f3}, we have made use of Equation \eqref{dm1}. From Equation \eqref{f3} to Equation \eqref{f4}, we have made use of $\text{Tr}[\partial_{\lambda}\rho]=0$ again. In the last step,
we have take advantage of Equation \eqref{internal}. We thus have the relation between first derivative of free energy and the reduced density matrix,
\begin{eqnarray}\label{central1}
\frac{\partial F}{\partial\lambda}=\sum_{ij}\text{Tr}\left[\frac{\partial\mathcal{U}^{ij}}{\partial\lambda}\rho^{ij}\right].
\end{eqnarray}
Differentiating both sides of Equation \eqref{central1} with respect to $\lambda$, we have
\begin{eqnarray}\label{central2}
\frac{\partial^2 F}{\partial\lambda^2}&=&\sum_{ij}\text{Tr}\left[\frac{\partial^2\mathcal{U}^{ij}}{\partial\lambda^2}\rho^{ij}\right]
+\sum_{ij}\text{Tr}\left[\frac{\partial\mathcal{U}^{ij}}{\partial\lambda}\frac{\partial\rho^{ij}}{\partial\lambda}\right].
\end{eqnarray}
From Equation \eqref{internal}, the first derivative of the free energy with respect to the temperature satisfies that
\begin{eqnarray}\label{central3}
F-T\frac{\partial F}{\partial T}&=&\sum_{ij}\text{Tr}[\mathcal{U}^{ij}\rho^{ij}].
\end{eqnarray}
Differentiating both sides of Equation \eqref{central3} with respect to temperature $T$, we get
\begin{eqnarray}\label{central4}
\frac{\partial^2 F}{\partial T^2}&=&-\frac{1}{T}\sum_{ij}\text{Tr}\left[\mathcal{U}^{ij}\frac{\partial\rho^{ij}}{\partial T}\right].
\end{eqnarray}
Equation \eqref{central1},\eqref{central2}, \eqref{central3} and \eqref{central4} are the first central results of the paper. We now make several comments on their implications to the phase transitions:\\
1. Equation \eqref{central1},\eqref{central2}, \eqref{central3} and \eqref{central4} connect the macroscopic quantities of a thermodynamic equilibrium state, the free energy and its derivatives, to the microscopic state of the system, two-body reduced density matrix of the canonical density matrix. In addition, Equation\eqref{central1} and \eqref{central2} recovers the results in \cite{Wu2004} at zero temperature. \\
2. Implications for first order phase transitions: First order phase transitions usually mean that the free energy is analytic function of the control parameters, such as $\lambda, T$, but the first derivatives of the free energy with respect to $\lambda, T$ are nonanalytic functions of the control parameters (nonanalytic may be either diverge or discontinuous). If we assume that $\mathcal{E}_{\alpha\beta}^i$ and $\mathcal{V}_{\alpha\beta\gamma\delta}^{i,j}$ are smooth functions of the control parameters of the system $\lambda$. From Equation \eqref{central1} and \eqref{central3}, the nonanalytic behavior of the first derivative of the free energy with respect to $\lambda, T$ must come from the nonanalytic behavior of matrix elements of the two-body reduced density matrix of the canonical density matrix, $\rho^{ij}$.\\
3. Implications for second order phase transitions: Second order phase transitions mean that the free energy and its first derivatives are analytic functions of the control parameters $\lambda, T$ but the second derivatives of the free energy with respect to the control parameters $\lambda, T$ are nonanalytic (either diverge or discontinuous). We assume that $\mathcal{E}_{\alpha\beta}^i$ and $\mathcal{V}_{\alpha\beta\gamma\delta}^{i,j}$ are smooth functions of the control parameters of the system $\lambda$. From Equation \eqref{central2} and \eqref{central4}, the nonanalytic behavior of the second derivatives of the free energy must come from the nonanalytic behavior of matrix elements of the first derivatives of the two-body reduced density matrix with respect to $\lambda, T$, i.e. $\partial_{\lambda}\rho^{ij}$ and $\partial_{T}\rho^{ij}$. Near thermal phase transitions, the singular part of the free energy $F_s$ presents the scaling behavior \cite{Cardy1996},
\begin{eqnarray}
F_s\left(\frac{1}{N},\delta T\right)=\Psi_0\left(\frac{b}{N},b^{1/\nu}\delta T\right).
\end{eqnarray}
Here $b$ is a scaling factor and $\delta T=T-T_c$ with $T_c$ being the critical temperature and $\delta\lambda=\lambda-\lambda_c$ with $\lambda_c$ being the critical control parameter. $\nu$ is the correlation length critical exponent of the thermal phase transitions and $\Psi_0(x,y)$ is a universal scaling function. Because Equation \eqref{central4} tells us that the singular part of the free energy must come from the singularity of the two-body reduced density matrix, we then expect that one of the matrix elements of the two-body reduced density matrix of the canonical density matrix near critical point satisfies the following scaling relation,
\begin{eqnarray}\label{scaling1}
\frac{\partial\rho^{ij}}{\partial T}\propto\Psi_1\left(\frac{b}{N},b^{1/\nu}\delta T\right).
\end{eqnarray}
Here $\Psi_1(x,y)$ is a universal scaling function. In the quantum critical region, the free energy density presents the scaling behavior \cite{Sachdev2011}
\begin{eqnarray}
F_s\left(\frac{1}{N},T,\delta\lambda\right)=\Psi_2\left(\frac{b}{N},b^{z}T,b^{1/\nu}\delta\lambda\right).
\end{eqnarray}
Here $\nu$ and $z$ are respectively the correlation length critical exponent and the dynamical critical exponent of the quantum phase transitions and $\Psi_2(x,y,z)$ is a universal scaling function. Because Equation \eqref{central2} tells us that the singular part of the free energy must come from the singularity of the first derivative of the two-body reduced density matrix, we then expect that one of the matrix elements of the two-body reduced density matrix of the canonical density matrix near critical point satisfies the following scaling relation,
\begin{eqnarray}\label{scaling2}
\frac{\partial\rho^{ij}}{\partial \lambda}\propto\Psi_3\left(\frac{b}{N},b^{z}T,b^{1/\nu}\delta\lambda\right).
\end{eqnarray}
Equation \eqref{scaling1} and \eqref{scaling2} tell us that the first derivative of the matrix elements of the two-body reduced density matrix present scaling behaviors both at quantum critical point and thermal critical point.

\section{Phase Transitions and Entanglement}
In recently years, entanglement measures have been used to diagnostic universal behaviors in quantum many-body systems, in particular phase transitions \cite{entanglementQPT2002,entanglementQPT2008,Renyientanglement2010}. The most useful entanglement measures are the R\'{e}nyi entropies and the von Neumann entropy \cite{entanglementQPT2002,entanglementQPT2008,Renyientanglement2010}. If a quantum system is prepared in a state $\rho$ and a bipartition of the system into a subsystem $A$ and its complement $B$, the reduced density matrix of part $A$ is $\rho_A=\text{Tr}_B[\rho]$. The R\'{e}nyi entropies $S_n$ of part $A$ are defined as \cite{Renyientanglement2010},
\begin{eqnarray}
S_A^{(n)}=\frac{1}{1-n}\ln\text{Tr}[\rho_A^n].
\end{eqnarray}
When $n\rightarrow1$, the Renyi entropy becomes von Neumann entropy, $\lim_{n\rightarrow1}S_A^{(n)}=S_A=-\text{Tr}[\rho_A\ln\rho_A]$.

In the previous section, we have established the connections between phase transitions and the reduced density matrix. Because quantum entanglement measures are defined from the reduced density matrix \cite{entanglementQPT2002,entanglementQPT2008,Renyientanglement2010}, it is thus conceivable that entanglement and phase transitions are directly connected. Now We first state the central theorem about phase transitions and entanglement at arbitrary temperatures, which is a generalization of the work by Wu and his collaborators \cite{Wu2004} to finite temperatures. Then we make a proof of the theorem.\\
\textbf{Theorem:} If the following conditions (i), (ii), (iii) are satisfied, then nonanalytic behavior in the R\'{e}nyi entanglement entropy and in the first derivative of the R\'{e}nyi entanglement entropy
are respectively necessary and sufficient conditions to signal a first order phase transition and a second order phase transitions.
\begin{enumerate}[label=(\roman*)]
\item The first order phase transition and second order phase transitions are associated to nonanalytic behavior of the first order derivative of the free energy and second order derivative of the free energy respectively.
Furthermore, the nonanalytic behavior in the first order derivative of the free energy and in the second order derivative of the free energy exclusively originated from the elements of the $\rho^{ij}$ and
not from the summation itself.
\item The nonanalytic matrix elements of $\rho^{ij}$ and its first derivatives $(\partial_{\lambda}\rho^{ij},\partial_T\rho^{ij})$ appear in the expression of R\'{e}nyi entanglement entropy do not either all accidentally vanish or cancel each other;
\item The nonanalytic matrix elements of $\rho^{ij}$ and its first derivatives $(\partial_{\lambda}\rho^{ij},\partial_T\rho^{ij})$ appear in the expression of R\'{e}nyi entanglement entropy do not either all accidentally vanish or cancel other terms in the expression for the four equations \eqref{central1},\eqref{central2}, \eqref{central3} and \eqref{central4}.
\end{enumerate}
Now let us prove the above theorem:\\
Proof: \textbf{The case for first order phase transitions:} If condition (i) is satisfied, then the first order phase transitions must come from nonanalytic behavior of one matrix elements of $\rho^{ij}$, as given by Equations \eqref{central1} and Equation \eqref{central3}. Taking the condition (ii)
into account, the first order phase transitions will be associated to nonanalytic behavior in the R\'{e}nyi entanglement entropy. So nonanalytic behavior in the R\'{e}nyi entanglement entropy is a necessary condition for first order phase transitions. Considering condition (iii), nonanalytic behavior in the the R\'{e}nyi entanglement entropy must come from the nonanalytic behavior of one or more of the matrix elements of the reduced density matrix $\rho^{ij}$. Assuming condition (i), a first order phase transitions follows. Thus nonanalytic behavior in the R\'{e}nyi entanglement entropy is also a sufficient condition for first order phase transitions.

\textbf{The case for second order phase transitions:} If condition (i) is satisfied, then the second order phase transitions must come from nonanalytic behavior of one or more of the matrix elements of the first derivative of the reduced density matrix $(\partial_T\rho^{ij},\partial_{\lambda}\rho^{ij})$, as given by Equations \eqref{central2} and Equation \eqref{central4}. Taking the condition (ii)
into account, the second order phase transitions will be associated to nonanalytic behavior in the first derivative of R\'{e}nyi entanglement entropy.
So nonanalytic behavior in the first derivative of R\'{e}nyi entanglement entropy is a necessary condition for second order
phase transitions. Considering condition (iii), the nonanalytic behavior in the first derivative of R\'{e}nyi entanglement entropy must come from the nonanalytic behavior of one or more of the matrix elements of the first derivative of the reduced density matrix, $(\partial_T\rho^{ij},\partial_{\lambda}\rho^{ij})$. Assuming condition (i), a second order phase transitions follows. Thus nonanalytic behavior in the first derivative of the R\'{e}nyi entanglement entropy is also a sufficient condition for second order phase transitions. Therefore the theorem is proved.

\begin{figure}
\begin{center}
\includegraphics[width=\columnwidth]{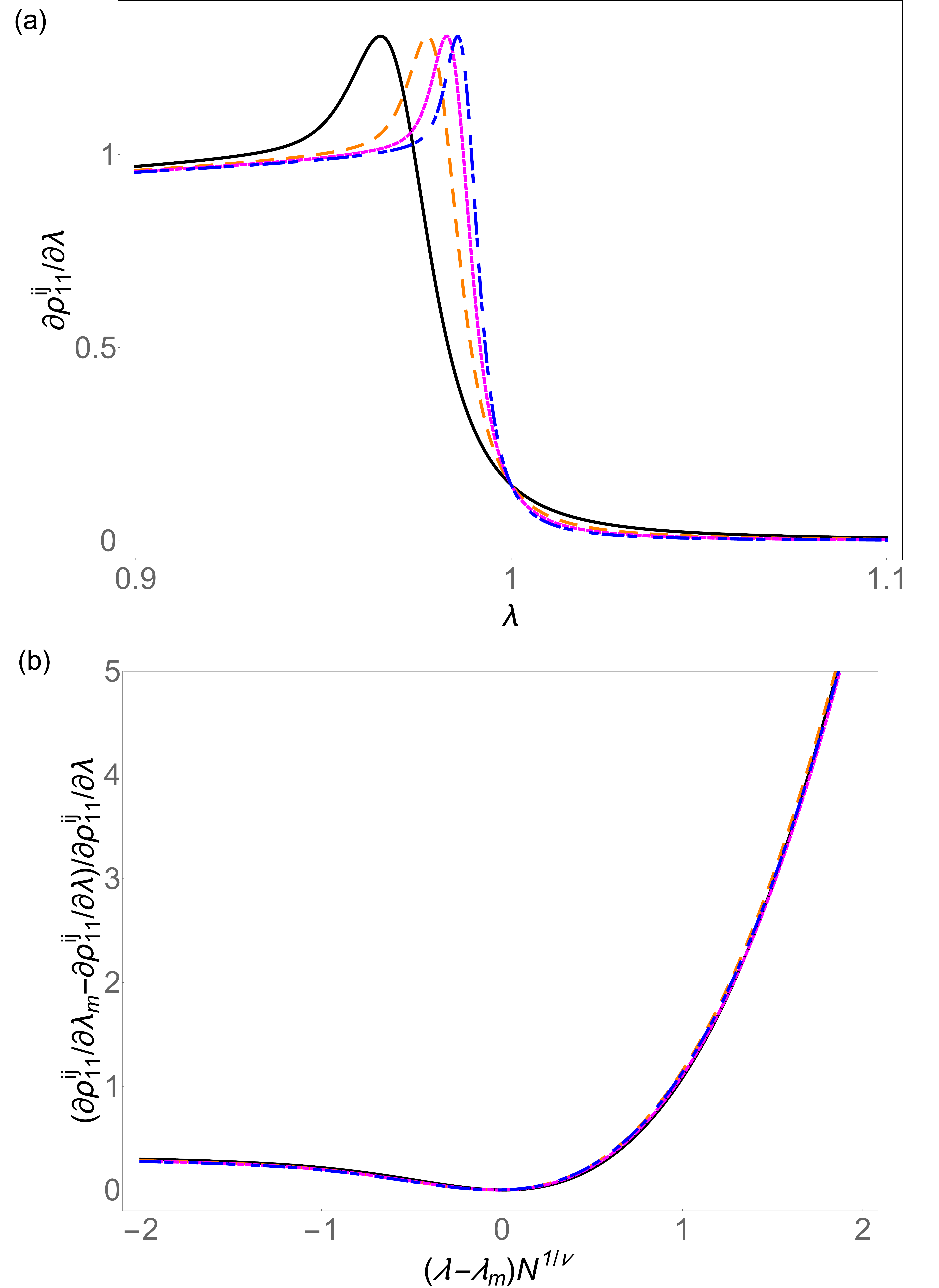}
\end{center}
\caption{(color online). Finite-size scaling of the first derivative of the two-body reduced density matrix element close to the quantum phase transition point in the LMG model. (a). The first derivative of the two-body reduced density matrix element with respect to control parameter $\lambda$, $\partial\rho_{11}^{ij}/\partial\lambda$, in the LMG model as a function of control parameter $\lambda$ for different number of spins $N$. The black solid line is $N=500$, the orange dashed line is $N=1000$, the magenta dotted line is $N=1500$ and the blue dash-dotted line is $N=2000$. (b). Data collapse of the first derivative of the two-body reduced density matrix element in the LMG model shown in (a). According to scaling arguments, $\partial\rho_{11}^{ij}/\partial\lambda$ is a function of $N^{1/\nu}(\lambda-\lambda_m)$ only with $\lambda_m$ being the position of the maximum of $\partial\rho_{11}^{ij}/\partial\lambda$ and $\nu=1.49$ being chosen so that data in (a) for different $N$ collapse perfectly. $\partial\rho_{11}^{ij}/\partial \lambda_m$ is a shorthand notation for $\partial\rho_{11}^{ij}/\partial \lambda|_{\lambda=\lambda_m}$.}
\label{fig:epsart1}
\end{figure}

\section{Physical Model Demonstration}
To demonstrate the above ideas, we study a many-body spin model with both quantum phase transitions and finite temperature phase transitions, namely the Lipkin-Meshkov-Glick (LMG) model \cite{LMG1965a,LMG1965b,LMG1965c} and the Hamiltonian of the LMG model is
\begin{eqnarray}
H=-\frac{J}{N}\sum_{i<j}\left(\sigma_i^x\sigma_j^x+\gamma\sigma_i^y\sigma_j^y\right)-\lambda\sum_j\sigma_j^z.
\end{eqnarray}
Here $J$ is the ferromagnetic coupling strength between two pauli spins $\vec{\sigma}_i$ and $\vec{\sigma}_j$ at arbitrary two sites along the $x$ and $y$ directions, $\gamma$ is the anisotropy of the ferromagnetic coupling in the $y$ direction, $\lambda$ is the magnetic field along $z$ direction. The LMG model and its various extensions have been experimentally realized in trapped ion systems \cite{Cirac2004,Fried2008,LMGExp2011} and also may be implemented in the nitrogen-vacancy centers system \cite{Wei2015EPJ}. Thus investigations in this work could be verified experimentally in near future.

Let us now relate the derivative of free energy and the matrix elements of the two-body reduced density matrices of the canonical density matrix:\\
\textbf{1. Free energy and its derivatives with respect to the control parameter $\lambda$:}
First, one can show that the diagonal matrix elements of the two-body reduced density matrix and the average value of physical quantity in the LMG model are related by (See Appendix for detailed derivations)
\begin{eqnarray}
\rho_{11}^{ij}&=&\frac{1}{4}\left[\langle\sigma_i^z\sigma_j^z\rangle+2\langle\sigma_j^z\rangle+1\right],\label{L1}\\
\rho_{44}^{ij}&=&\frac{1}{4}\left[\langle\sigma_i^z\sigma_j^z\rangle-2\langle\sigma_j^z\rangle+1\right],\label{L2}\\
\rho_{22}^{ij}&=&\rho_{33}^{ij}=\frac{1}{4}\left(1-\langle\sigma_i^z\sigma_j^z\rangle\right).\label{L3}
\end{eqnarray}
Then the first derivative of the free energy can be calculated as
\begin{eqnarray}
\frac{\partial F}{\partial\lambda}&=&\left\langle\frac{\partial H}{\partial\lambda}\right\rangle,\\
&=&-N\langle\sigma_j^z\rangle,\\
&=&-N\left[\rho_{11}^{ij}-\rho_{44}^{ij}\right].
\end{eqnarray}
Here we have made use of the translation symmetry and Equations \eqref{L1} to \eqref{L3}. Differentiating the above equation with respect to $\lambda$, we get
\begin{eqnarray}\label{F2}
\frac{\partial^2 F}{\partial\lambda^2}&=&-N\left[\frac{\partial\rho_{11}^{ij}}{\partial\lambda}-\frac{\partial\rho_{44}^{ij}}{\partial\lambda}\right].
\end{eqnarray}
We thus proved analytically Equations \eqref{central1} and \eqref{central2} in the LMG model. Because the LMG model presents a second order quantum phase transitions from a ferromagnetic phase to a paramagnetic phase at critical field $\lambda_c=1$, we thus expect that $\frac{\partial\rho_{11}^{ij}}{\partial\lambda}$ or $\frac{\partial\rho_{44}^{ij}}{\partial\lambda}$ presents universal scaling behavior.

In Figure 1, we show that the critical behavior of the first derivative of the two-body reduced density matrix element $\partial\rho^{11}/\partial\lambda$ of the ground state as a function of the control parameter $\lambda$.
In Figure 1 (a), we plot $\partial\rho^{11}/\partial\lambda$ as a function of control parameter $\lambda$ for the system with different number of spins $N=500,1000,1500,2000$ respectively. First, one can see that $\partial\rho^{11}/\partial\lambda$ for systems with different number of spins cross at the quantum critical point $\lambda_c=1$. Second, $\partial\rho^{11}/\partial\lambda$ presents a peak at $\lambda_m$ which is close to the critical point. As the system size increases, the position of control parameter $\lambda_m$ where $\partial\rho^{11}/\partial\lambda$ has a peak approaches the quantum critical point $\lambda_c$. In Figure 1(b), we plot the $(\partial\rho^{11}/\partial\lambda|_{\lambda_m}-\partial\rho^{11}/\partial\lambda)/\partial\rho^{11}/\partial\lambda$ as a function of scale parameter $(\lambda-\lambda_m)N^{1/\nu}$. We choose $\nu$ so that the data in Figure 1(a) collapse perfectly and we found that the correlation length critical exponent $\nu=1.49$, which is close to the exact value $\nu=3/2$ \cite{LMG1983}.

\begin{figure}
\begin{center}
\includegraphics[width=\columnwidth]{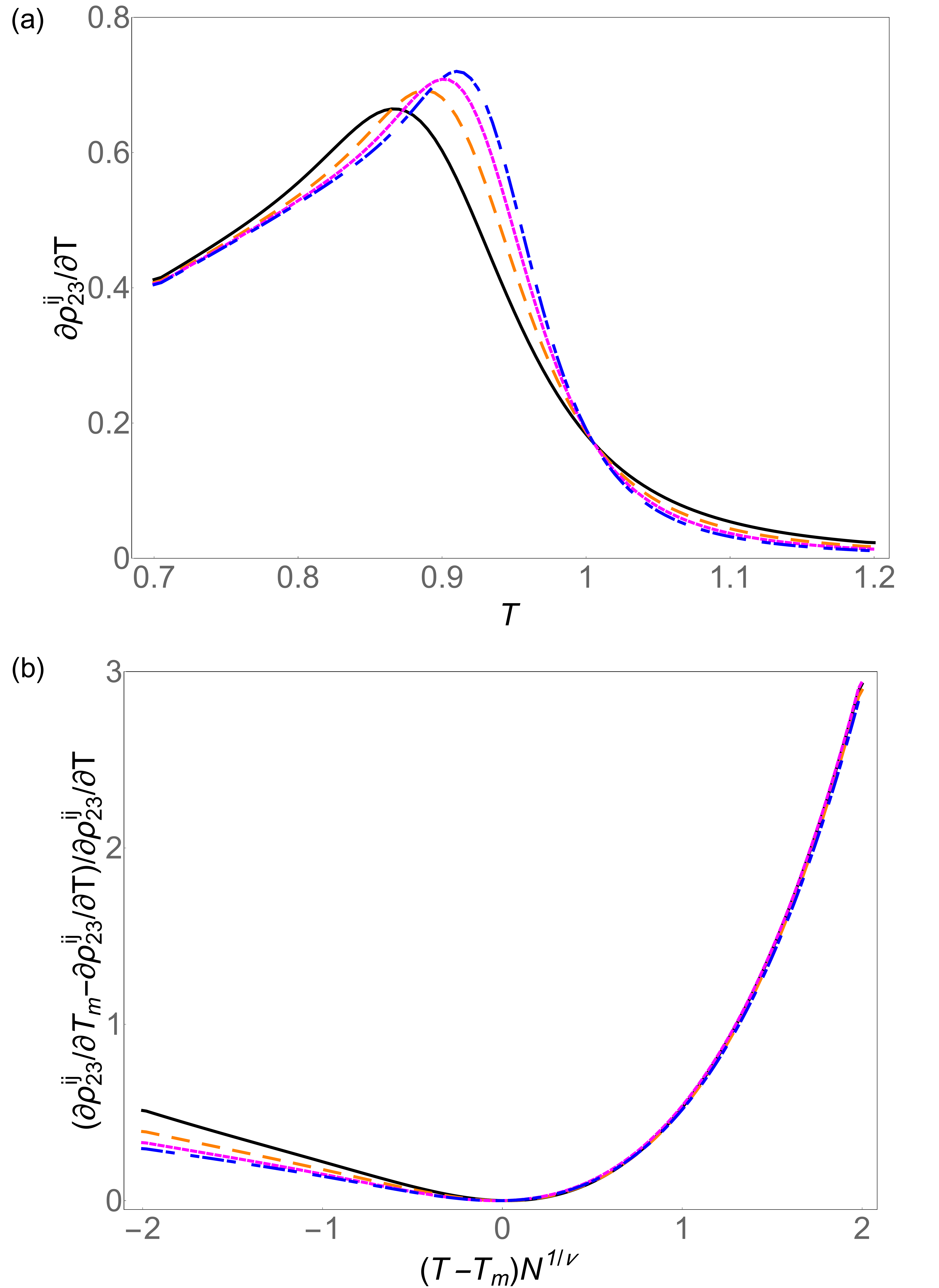}
\end{center}
\caption{(color online). Finite-size scaling of the first derivative of the two-body reduced density matrix close to the thermal phase transition point in the LMG model. (a). The first derivative of the two-body reduced density matrix element with respect to temperature, $\partial \rho_{23}^{ij}/\partial T$, in the LMG model as a function of temperature $T$ for different number of spins $N$. The black solid line is $N=200$, the orange dashed line is $N=300$, the magenta dotted line is $N=400$ and the blue dash-dotted line is $N=500$. Here we only show the real part of $\rho_{23}^{ij}$ because it is a complex number. (b). Data collapse of the first derivative of the two-body reduced density matrix element with respect to temperature in the LMG model shown in (a). According to scaling arguments, $\partial\rho_{23}^{ij}/\partial T$ is a function of $N^{1/\nu}(T-T_m)$ only with $T_m$ being the position of the maximum of $\partial\rho_{23}^{ij}/\partial T$ and $\nu=2.01$ being chosen so that data in (a) for different $N$ collapse perfectly. $\partial\rho_{23}^{ij}/\partial T_m$ is a shorthand notation for $\partial\rho_{23}^{ij}/\partial T|_{T=T_m}$.}
\label{fig:epsart2}
\end{figure}

\textbf{2. Free energy and its derivatives with respect to temperature $T$:} The off-diagonal matrix elements of the two-body reduced density matrix and the average values of the physical observable are related by (See Appendix for detailed derivations)
\begin{eqnarray}
\Re\rho_{23}^{ij}&=&\frac{1}{4}\left[\langle\sigma_i^x\sigma_j^x\rangle+\langle\sigma_i^y\sigma_j^y\rangle\right],\label{L4}\\
\Re\rho_{14}^{ij}&=&\frac{1}{4}\left[\langle\sigma_i^x\sigma_j^x\rangle-\langle\sigma_i^y\sigma_j^y\rangle\right].\label{L5}
\end{eqnarray}
The average value of the internal energy is
\begin{eqnarray}
\langle E\rangle&=&\langle H\rangle,\\
&=&-\frac{J(N-1)}{2}\left[\langle\sigma_i^x\sigma_j^x\rangle+\gamma\langle\sigma_i^y\sigma_j^y\rangle\right]-\lambda N\langle\sigma_j^z\rangle,\\
&=&-(N-1)J\left[(1-\gamma)\Re\rho_{14}^{ij}+(1+\gamma)\Re\rho_{23}^{ij}\right]\nonumber\\
&& -\lambda N\left(\rho_{11}^{ij}-\rho_{44}^{ij}\right),\\
&=&F-T\frac{\partial F}{\partial T}.
\end{eqnarray}
In the above derivations, we have made use of the translation symmetry and Equations \eqref{L4} and \eqref{L5}. Differentiating the above equation with respect to temperature, we get
\begin{eqnarray}
\frac{\partial^2 F}{\partial T^2}&=&\frac{(N-1)J}{T}\left[(1-\gamma)\frac{\partial\Re\rho_{14}^{ij}}{\partial T}+(1+\gamma)\frac{\partial\Re\rho_{23}^{ij}}{\partial T}\right]\nonumber\\
&& +\frac{\lambda N}{T}\left[\frac{\partial\rho_{11}^{ij}}{\partial T}-\frac{\partial\rho_{44}^{ij}}{\partial T}\right].
\end{eqnarray}
We thus proved analytically Equations \eqref{central2} and \eqref{central4} in the LMG model.

In Figure 2, we show that the critical behavior of the first derivative of the two-body reduced density matrix element $\partial\rho^{11}/\partial T$ of the equilibrium state as a function of temperature $T$.
In Figure 2 (a), we plot $\partial\rho^{ij}_{23}/\partial T$ as a function of temperature $T$ for the system with different number of spins $N=200,300,400,500$ respectively. First, one can see that $\partial\rho^{ij}_{23}/\partial T$ for systems with different number of spins cross at the thermal critical point $T_c=1$. Second, $\partial\rho^{ij}_{23}/\partial T$ presents a peak at $T_m$ which is close to the thermal critical point. As the system size increases, the position of temperature $T_m$ where $\partial\rho^{ij}_{23}/\partial T$ has a peak approaches the thermal critical point $T_c$. In Figure 2(b), we plot the $(\partial\rho^{ij}_{23}/\partial T|_{T_m}-\partial\rho^{ij}_{23}/\partial T)/\partial\rho^{ij}_{23}/\partial T$ as a function of scale variable $(T-T_m)N^{1/\nu}$. We choose $\nu$ so that the data in Figure 2(a) collapse perfectly and we found that the correlation length critical exponent $\nu=2.01$, which is close to the exact value $\nu=2$ \cite{LMG1983}.

\section{Summary}
In summary, we have established a general theory of phase transitions and quantum entanglement in the equilibrium state at arbitrary temperature. We derived a set of universal functional relations between the matrix elements of two-body reduced density matrix of the canonical equilibrium state and the Helmholtz free energy of the equilibrium state. These relations imply that the free energy and its derivatives are directly related to quantum entanglement in the canonical equilibrium state. Furthermore, we showed that the first order phase transitions are signaled by the matrix elements of reduced density matrix while the second order phase transitions are witnessed by the first derivatives of the reduced density matrix elements. Close to second order phase transitions, we showed that the first derivatives of the reduced density matrix elements present universal scaling behaviors. We finally established a theorem which connects the phase transitions and entanglement at arbitrary temperature. Our general results are demonstrated in the LMG model and could be verified experimentally in trapped ion settings.

\begin{acknowledgements}
This work was supported by the National Natural Science
Foundation of China (Grant Number 11604220).
\end{acknowledgements}

\section*{Appendix: Derivation of the Relations between the Matrix Elements of Two-body Reduced Density Matrix and the Average Value of Physical Quantity in LMG Model}
 \renewcommand{\theequation}{A\arabic{equation}} \setcounter{equation}{0}
In this appendix we derive the relations between the matrix elements of two-body reduced density matrix and the average value of physical quantity in LMG model.
First the average of a Pauli spin along $z$ direction can be calculated as
\begin{eqnarray}
\langle\sigma_j^z\rangle&=&\text{Tr}[\rho\sigma_j^z],\\
&=&\text{Tr}_{ij}[\rho^{ij}\sigma_j^z],\\
&=&\sum_{\alpha,\beta}\langle\alpha_i\beta_j|\rho^{ij}\sigma_j^z|\alpha_i\beta_j\rangle,\\
&=&\rho_{11}^{ij}-\rho_{22}^{ij}+\rho_{33}^{ij}-\rho_{44}^{ij}.
\end{eqnarray}
Similarly, one gets
\begin{eqnarray}
\langle\sigma_i^z\rangle&=&\text{Tr}[\rho\sigma_i^z],\\
&=&\rho_{11}^{ij}+\rho_{22}^{ij}-\rho_{33}^{ij}-\rho_{44}^{ij}.
\end{eqnarray}
Translation symmetry implies that $\langle\sigma_i^z\rangle=\langle\sigma_j^z\rangle$, which leads to
\begin{eqnarray}
\rho_{22}^{ij}=\rho_{33}^{ij}.
\end{eqnarray}
Thus
\begin{eqnarray}
\langle\sigma_j^z\rangle&=&\rho_{11}^{ij}-\rho_{44}^{ij}.
\end{eqnarray}
Besides, $\text{Tr}[\rho]=1$ tells us that
\begin{eqnarray}
\rho_{11}^{ij}+2\rho_{22}^{ij}+\rho_{44}^{ij}=1.
\end{eqnarray}
The correlation function of two Pauli spins along $z$ direction is
\begin{eqnarray}
\langle\sigma_i^z\sigma_j^z\rangle&=&\text{Tr}[\rho\sigma_i^z\sigma_j^z],\\
&=&\rho_{11}^{ij}-\rho_{22}^{ij}-\rho_{33}^{ij}+\rho_{44}^{ij},\\
&=&\rho_{11}^{ij}-2\rho_{22}^{ij}+\rho_{44}^{ij},\\
&=&2[\rho_{11}^{ij}+\rho_{44}^{ij}]-1.
\end{eqnarray}
Therefore, we have
\begin{eqnarray}
\rho_{11}^{ij}&=&\frac{1}{4}\left[\langle\sigma_i^z\sigma_j^z\rangle+2\langle\sigma_j^z\rangle+1\right],\\
\rho_{44}^{ij}&=&\frac{1}{4}\left[\langle\sigma_i^z\sigma_j^z\rangle-2\langle\sigma_j^z\rangle+1\right],\\
\rho_{22}^{ij}&=&\rho_{33}^{ij}=\frac{1}{4}\left(1-\langle\sigma_i^z\sigma_j^z\rangle\right).
\end{eqnarray}
Thus Equations \eqref{L1}, \eqref{L2} and Equation \eqref{L3} in the main text are derived.

One can see that the Hamiltonian of LMG model is invariant under a global rotation along $z$ axis by an angle $\pi$. This leads to
\begin{eqnarray}
\langle \sigma_i^x\rangle&=&0,\\
\langle \sigma_i^y\rangle&=&0.
\end{eqnarray}
The average value of $\sigma_i^x$ and $\sigma_i^y$ can be given by the matrix elements of the two-body reduced density matrix,
\begin{eqnarray}
\langle\sigma_j^x\rangle&=&\text{Tr}[\rho\sigma_j^x],\\
&=&\text{Tr}_{ij}[\rho^{ij}\sigma_j^x],\\
&=&\sum_{\alpha,\beta}\langle\alpha_i\beta_j|\rho^{ij}\sigma_j^x|\alpha_i\beta_j\rangle,\\
&=&\sum_{\alpha,\beta,\gamma,\delta}\langle\alpha_i\beta_j|\rho^{ij}|\gamma_i\delta_j\rangle\langle\gamma_i\delta_j|\sigma_j^x|\alpha_i\beta_j\rangle,\\
&=&\rho_{12}^{ij}+\rho_{21}^{ij}+\rho_{34}^{ij}+\rho_{43}^{ij},\\
&=&2\left[\Re\rho_{12}^{ij}+\Re\rho_{34}^{ij}\right].
\end{eqnarray}
\begin{eqnarray}
\langle\sigma_j^y\rangle&=&\text{Tr}[\rho\sigma_j^y],\\
&=&\text{Tr}_{ij}[\rho^{ij}\sigma_j^y],\\
&=&\sum_{\alpha,\beta}\langle\alpha_i\beta_j|\rho^{ij}\sigma_j^y|\alpha_i\beta_j\rangle,\\
&=&\sum_{\alpha,\beta,\gamma,\delta}\langle\alpha_i\beta_j|\rho^{ij}|\gamma_i\delta_j\rangle\langle\gamma_i\delta_j|\sigma_j^y|\alpha_i\beta_j\rangle,\\
&=&-i\rho_{12}^{ij}+i\rho_{21}^{ij}-i\rho_{34}^{ij}+i\rho_{43}^{ij},\\
&=&2\left[\Im\rho_{12}^{ij}+\Im\rho_{34}^{ij}\right].
\end{eqnarray}
Thus we have
\begin{eqnarray}
\rho_{12}^{ij}=-\rho_{34}^{ij}.
\end{eqnarray}
The average value of two-body operators can be calculated as
\begin{eqnarray}
\langle\sigma_i^x\sigma_j^x\rangle&=&\text{Tr}[\rho\sigma_i^x\sigma_j^x],\\
&=&\text{Tr}_{ij}[\rho^{ij}\sigma_i^x\sigma_j^x],\\
&=&\sum_{\alpha,\beta}\langle\alpha_i\beta_j|\rho^{ij}\sigma_i^x\sigma_j^x|\alpha_i\beta_j\rangle,\\
&=&\sum_{\alpha,\beta,\gamma,\delta}\langle\alpha_i\beta_j|\rho^{ij}|\gamma_i\delta_j\rangle\langle\gamma_i\delta_j|\sigma_i^x\sigma_j^x|\alpha_i\beta_j\rangle,\\
&=&\rho_{14}^{ij}+\rho_{23}^{ij}+\rho_{32}^{ij}+\rho_{41}^{ij},\\
&=&2\left[\Re\rho_{14}^{ij}+\Re\rho_{23}^{ij}\right].
\end{eqnarray}
In the above, we have made use of the Hermitian property of the two-body reduced density matrix. Moreover,
\begin{eqnarray}
\langle\sigma_i^y\sigma_j^y\rangle&=&\text{Tr}[\rho\sigma_i^y\sigma_j^y],\\
&=&\text{Tr}_{ij}[\rho^{ij}\sigma_i^y\sigma_j^y],\\
&=&\sum_{\alpha,\beta}\langle\alpha_i\beta_j|\rho^{ij}\sigma_i^y\sigma_j^y|\alpha_i\beta_j\rangle,\\
&=&\sum_{\alpha,\beta,\gamma,\delta}\langle\alpha_i\beta_j|\rho^{ij}|\gamma_i\delta_j\rangle\langle\gamma_i\delta_j|\sigma_i^y\sigma_j^y|\alpha_i\beta_j\rangle,\\
&=&-\rho_{14}^{ij}+\rho_{23}^{ij}+\rho_{32}^{ij}-\rho_{41}^{ij},\\
&=&2\left[\Re\rho_{23}^{ij}-\Re\rho_{14}^{ij}\right].
\end{eqnarray}
We thus obtain
\begin{eqnarray}
\Re\rho_{23}^{ij}&=&\frac{1}{4}\left[\langle\sigma_i^x\sigma_j^x\rangle+\langle\sigma_i^y\sigma_j^y\rangle\right],\\
\Re\rho_{14}^{ij}&=&\frac{1}{4}\left[\langle\sigma_i^x\sigma_j^x\rangle-\langle\sigma_i^y\sigma_j^y\rangle\right].
\end{eqnarray}
Thus Equations \eqref{L4} and Equation \eqref{L5} in the main text are derived.

\end{document}